\documentclass[aps,prl,twocolumn,showpacs,amsmath,amssymb,amsfonts,nofootinbib,long]{revtex4}
\usepackage{graphicx}
\begin{document}
\title{Scattering of Dirac and Majorana Fermions off Domain Walls}
\author{Leonardo Campanelli$^{1,2}$}
\email{campanelli@fe.infn.it}
\affiliation{$^{1}${\it Dipartimento di Fisica,
             Universit\`a di Ferrara, I-44100 Ferrara, Italy \\
             $^{2}$INFN - Sezione di Ferrara, I-44100 Ferrara, Italy}}
\date{September, 2004}


\begin{abstract}
We investigate the interaction of fermions having both Dirac and
left-handed and right-handed Majorana mass terms with vacuum
domain walls. By solving the equations of motion in thin-wall
approximation, we calculate the reflection and transmission
coefficients for the scattering of fermions off walls.
\end{abstract}

\pacs{11.27.+d, 14.60.Pq}
\maketitle


\section{I. Introduction}
Theoretical arguments \cite{Friedland} and a recent analysis of
WMAP data \cite{Conversi} show that the presence of a network of
low-tension domain walls in the Universe is not ruled out.
Moreover, domain walls could provide a natural and non-exotic
alternative to the most popular candidates of dark energy
\cite{Sahni}. The evolution of vacuum domain walls in the early
Universe is determined by their interaction with the surrounding
plasma. The two most relevant effects to be considered are the
particle scattering off walls, and the presence of bound states
near the walls, the so-called ``zero modes''.

The scattering of particles off walls (including the scattering
between walls) determines the average velocity $v$ of a wall and
thus, in turn, the equation of state of a gas of domain walls,
$p_w = (v^2 - 2/3) \rho_w$, where $\rho_w$ and $p_w$ are the
energy density and pressure of the gas (see, e.g., \cite{KOLB}).
Indeed, when particles scatter off a wall, they generate a
frictional force $F = \sum_i n_i R_i \Delta p_i$, where $n_i$ is
the number density of particles of species $i$, $R_i$ is their
scattering probability (the reflection coefficient), and $\Delta
p_i$ is the momentum transfer per collision (see, e.g.,
\cite{VILENKIN}). Hence, defining the mean velocity of the walls
as $v = \sum_i n_i R_i v_i / \sum_i n_i R_i$, the damping force
can be written as $F = \mu v $, where we have defined the
frictional coefficient $\mu = (\sum_i n_i R_i \Delta p_i)(\sum_i
n_i v_i) / \sum_i n_i R_i$. The mean velocity of a wall is
determined by balancing the tension, $f \sim \sigma /r$, where
$\sigma$ and $r$ are the surface energy density and the mean
curvature radius of a wall, and the friction, $f = F$. The
resulting velocity is then $v \sim \sigma / r \mu$. It is clear
that a full analysis of the role of domain walls in the Universe
imposes the study of their interaction with particles in the
primordial plasma.

The presence of zero modes localized on a domain wall can be
important for the stability of the wall. In particular, fermionic
zero modes may give rise to interesting phenomena as the
magnetization of domain walls \cite{Iwazaki,Cea1}, and the
dynamical generation of massive ferromagnetic domain walls
\cite{Campanelli1}. Indeed, fermionic zero modes could drastically
change both gravitational properties and cosmic evolution of a gas
of domain walls \cite{Campanelli2}.

The interaction of scalar particles and Dirac fermions with a
domain wall has been the object of various papers in the
literature (see \cite{VILENKIN} and references therein,
\cite{Voloshin,Jackiw,Callan,Farrar,Cea2,Campanelli3,Ayala}).
Since strong evidence for neutrino masses has emerged from various
neutrino oscillation experiments in recent years \cite{Grossman},
we are motivated to investigate the interaction of Majorana
fermions with domain walls (neutrinos are neutral fermions, and
then can have both Majorana and Dirac masses).
In a recent paper \cite{Stojkovic}, Stojkovic has studied
fermionic zero modes in the domain wall background, in the case in
which the fermions have both Dirac and left-handed and
right-handed Majorana mass terms. The aim of this paper is to
study the scattering of such fermions off domain walls.

The plan of the paper is as follows. In Section II we introduce
the Lagrangian for a single real self-interacting scalar field
$\Phi$, coupled with a fermion $\psi$ having Dirac, left-handed
and right-handed Majorana mass terms. We also derive the equations
of motion. In Section III we calculate the reflection and
transmission coefficients for the scattering of fermions off
walls, in both cases in which the coupling to the scalar field
$\Phi$ is or not the source of the Majorana mass terms. Finally,
we summarize our results in Section IV.


\section{II. Lagrangian, asymptotic states, and equations of motion}
We consider a simplified model in which the kink is a infinitely
static domain wall in the $xz$-plane. In this model the scalar
sector giving rise to a planar wall is a real scalar field with
density Lagrangian
\begin{equation}
\label{Eq1}
{\cal L}_{\Phi} = \frac{1}{2} \,
\partial_{\mu} \Phi\partial^{\mu} \Phi
- \frac{\lambda}{4} (\Phi^{2} - \eta^{2})^2.
\end{equation}
In the tree approximation, the set of vacuum states is $\langle
\Phi \rangle^2 = \eta^{2}$, so that one may assume that there are
regions with $\langle \Phi \rangle = + \eta$ and regions with
$\langle \Phi \rangle = - \eta$. By continuity there must exists a
region in which the scalar field is out of the vacuum. This region
is a domain wall \cite{Kibble}, and the classical equation of
motion admits the solution describing the transition layer between
two regions with different values of $\langle \Phi \rangle$,
\begin{equation}
\label{Eq2}
\Phi(y) = \eta \tanh (y / \Delta),
\end{equation}
where $\Delta = \sqrt{2 \lambda}/\eta$ is the thickness of the
wall \cite{KOLB}.
\\
Now, we consider a fermion $\psi$ having Dirac, left-handed and
right-handed Majorana mass terms. The source of the Dirac and
Majorana mass terms is the Yukawa couplings to the scalar field
$\Phi$. In terms of the chiral spinors $\psi_{L}$ and $\psi_{R}$
the Lagrangian density of the system is
\footnote{For a full discussion of validity of Lagrangian
(\ref{Eq3}) in phenomenologically relevant models in neutrino
physics, see Ref. \cite{Stojkovic} and references therein.}:
\begin{eqnarray}
\label{Eq3}
{\cal {L}} \!\!& = &\!\! \mbox{$\frac{1}{2}$} \,
\partial_{\mu} \Phi\partial^{\mu} \Phi -
\mbox{$\frac{\lambda}{4}$} (\Phi^2 - \eta^2)^2
\nonumber \\
\!\!& + &\!\! i \bar{\psi}_{L} \, / \!\!\!\partial \, \psi_{L} + i
\bar{\psi}_{R} \, / \!\!\!\partial \, \psi_{R} \\
\!\!& - &\!\! (g_D \Phi \bar{\psi}_{L} \psi_{R} + g_L \Phi
\bar{\psi}_{L} \psi_{L}^{c} + g_R \Phi \bar{\psi}_{R} \psi_{R}^{c}
+ h.c.), \nonumber
\end{eqnarray}
where $g_D$, $g_L$ and $g_R$ are the Yukawa couplings to the
scalar field of the Dirac, left-handed and right-handed Majorana
fermions, respectively. In Lagrangian (\ref{Eq3}), $\psi^{c} = C
\bar{\psi}^{T}$, where $C = i \gamma^{2} \gamma^{0}$ is the charge
conjugation matrix \cite{BJORKEN} and ``$\, T \,$'' indicates the
transpose.

In the broken phase (i.e. for $y \rightarrow \pm \infty$) where
$\Phi$ takes a constant value, $\langle \Phi \rangle = \pm \eta$,
the scalar field give mass to the fermions states. It is clear
that the chiral fields $\psi_L$ and $\psi_R$ do not have a
definite mass, since they are coupled by the Dirac mass term. In
order to find the asymptotic states with definite masses, we have
to diagonalise the mass matrix in Lagrangian (\ref{Eq3}) or,
equivalently, we have to diagonalize the Dirac equation,
\begin{equation}
\label{Eq22} (i \slash\!\!\!\partial - G \Phi) \Psi = 0,
\end{equation}
where we have introduced the following quantities:
\begin{equation}
\label{Eq23} G = \begin{pmatrix}
                     0 & \mathcal{G} \\
                     \mathcal{G} & 0
              \end{pmatrix} \! ,
\;\;\; \mathcal{G} = \begin{pmatrix}
                     g_L & g_D \\
                     g_D & g_R
              \end{pmatrix} \! ,
\;\;\; \Psi = \left( \!\!\!
                       \begin{array}{c}
                            \psi_L    \\
                            \psi_R^c  \\
                            \psi_L^c  \\
                            \psi_R
                       \end{array}
       \!\! \right) \! .
\end{equation}
In the broken phase, Eq.~(\ref{Eq22}) becomes:
\begin{equation}
\label{Eq24} (i \, \slash\!\!\!\partial \mp M) \Psi = 0, \;\;\;
\mbox{if $y \rightarrow \pm \infty$},
\end{equation}
where we have defined the ``mass matrix'' $M = \eta G$.
Diagonalizing the matrix $M$, we get:
\begin{equation}
\label{Eq25} (i \slash\!\!\!\partial \mp \Delta) \Psi_M = 0,
\;\;\; \mbox{if $y \rightarrow \pm \infty$},
\end{equation}
where
\begin{equation}
\label{Eq26} \Delta = U^T M U
       = \begin{pmatrix}
                     m_1 & 0 & 0 & 0  \\
                     0 & m_2 & 0 & 0  \\
                     0 & 0 & -m_1 & 0 \\
                     0 & 0 & 0 & -m_2
         \end{pmatrix} \!,
\end{equation}
$U$ is the unitary transformation which diagonalizes $M$, and
$\Psi_M = U^T \Psi$. The eigenvalues of $M$, that is $\pm
m_{1,2}$, are given by
\begin{equation}
\label{Eq12new} m_{1,2} = \frac{1}{2} \! \left [ m_L + m_R \pm
\sqrt{4 m_D^2 + (m_L - m_R)^2} \: \right ] \!\! ,
\end{equation}
where we have defined
\begin{eqnarray}
\label{Eq13new} m_D = g_D \eta ,  \;\;\; m_L = g_L \eta , \;\;\;
m_R = g_R \eta .
\end{eqnarray}
Here, $m_1$ and $m_2$ represent the masses of the free-field
propagating degrees of freedom in the theory. It can be showed
(see e.g. Ref.~\cite{Grossman}) that the two massive fermion
states are Majorana particles.

After having considered the asymptotic fermion states, it is now
clear that the particle content of Lagrangian (\ref{Eq3}) consist
of two Majorana fermions with masses $m_1$ and $m_2$ interacting
with a vacuum domain wall (described by the scalar field $\Phi$).
The aim of this paper is to study the scattering of this two
states off a wall. To this end, we use the following
representation of the Dirac matrices,
\begin{eqnarray}
\label{Eq4} \gamma^0 \!\!& = &\!\!
          \begin{pmatrix} \sigma^3 & 0 \\ 0 &
          -\sigma^3 \end{pmatrix}\!\!, \;\;
\gamma^1 =
          \begin{pmatrix} i\sigma^2 & 0 \\ 0 &
          -i\sigma^2 \end{pmatrix}\!\!, \nonumber \\
\gamma^2 \!\!& = &\!\!
          \begin{pmatrix} -i\sigma^1 & 0 \\ 0 &
          i\sigma^1 \end{pmatrix}\!\!, \;\;
\gamma^3 =
          \begin{pmatrix}0 & 1 \\ -1
          & 0 \end{pmatrix} \!\! ,
\end{eqnarray}
where $\sigma^k$, $k = 1,2,3$, are the Pauli matrices. In this
representation, a four-component fermion has left-handed and
right-handed component of the form
\begin{equation}
\label{Eq5} \psi_{L}^{T} = (\alpha,\beta,-\alpha,-\beta), \;\;
\psi_{R}^{T} = (\gamma,\delta,\gamma,\delta).
\end{equation}
We will concentrate on the solution describing the motion of
fermions perpendicular to the wall, i.e. along the $y$-axis, and
then we suppose that
\begin{equation}
\label{Eq6} \Phi = \Phi(y), \;\; \psi_{L} = \psi_{L}(y,t), \;\;
\psi_{R} = \psi_{R}(y,t).
\end{equation}
The Lagrangian (\ref{Eq3}), together with Eqs.~(\ref{Eq5}) and
(\ref{Eq6}), implies the equations of motion
\begin{equation}
\label{Eq7} \Phi'' - \lambda \Phi (\Phi^2 - \eta^2)
       = 4 g_D \mbox{Re} (\alpha^* \gamma - \beta^* \delta),
\end{equation}
and
\begin{eqnarray}
\label{Eq8} \beta'  + i \dot{\alpha}  \!\!& = &\!\!
                                g_D \Phi \gamma + g_L \Phi \beta^*,  \nonumber \\
\alpha' - i \dot{\beta}   \!\!& = &\!\!
                                g_D \Phi \delta + g_L \Phi \alpha^*, \nonumber \\
\delta' + i \dot{\gamma}  \!\!& = &\!\!
                                g_D \Phi \alpha + g_R \Phi \delta^*,           \\
\gamma' - i \dot{\delta}  \!\!& = &\!\!
                                g_D \Phi \beta + g_R \Phi \gamma^*, \nonumber
\end{eqnarray}
where $\text{Re}(x)$ is the real part of $x$ (here, and
throughout, a prime and a dot will denote differentiation with
respect to $y$ and $t$, respectively).

In the following we shall analyze the simple case in which the
back-reaction of the fermion field $\psi$ on the domain wall
configuration is null. Indeed, we make the ansatz: $\beta =
\alpha^*$ and $\gamma = \delta^*$, which is compatible with
Eq.~(\ref{Eq8}), and makes null the right hand side of
Eq.~(\ref{Eq7}). [This, in turns, means that the wall profile is
given by Eq.~(\ref{Eq2}).] Moreover, writing $\alpha$ and $\delta$
as a sum of positive and negative energy states,
\begin{eqnarray}
\label{Eq9} \alpha(y,t) \!\!& = &\!\!   \alpha_{+}(y) \, e^{-iEt}
                          + \alpha_{-}(y) \, e^{iEt}, \nonumber \\
\delta(y,t) \!\!& = &\!\!   \delta_{+}(y) \, e^{-iEt}
                          + \delta_{-}(y) \, e^{iEt},
\end{eqnarray}
and inserting into Eq.~(\ref{Eq8}) we get
\begin{eqnarray}
\label{Eq10} {\alpha^*_{-}}' + E \alpha_{+} \!\!& = &\!\!
                                     g_D \Phi \delta^*_{-} + g_L \Phi \alpha_{+}, \nonumber \\
{\delta^*_{-}}' - E \delta_{+} \!\!& = &\!\!
                                     g_D \Phi \alpha^*_{-} + g_R \Phi \delta_{+}, \nonumber \\
{\alpha^*_{+}}' - E \alpha_{-} \!\!& = &\!\!
                                     g_D \Phi \delta^*_{+} + g_L \Phi \alpha_{-},           \\
{\delta^*_{+}}' + E \delta_{-} \!\!& = &\!\!
                                     g_D \Phi \alpha^*_{+} + g_R \Phi \delta_{-}. \nonumber
\end{eqnarray}
Starting from Eq.~(\ref{Eq10}), we will calculate, in the next
Section, the reflection and transmission coefficients for the
scattering of fermions off walls.


\section{III. Scattering}

We will work in ``thin-wall approximation'' \cite{VILENKIN}, that
is to say we suppose that the thickness of the wall is vanishingly
small, $\Delta \rightarrow 0$. In this case, the wall profile
takes the simple form $\Phi = \eta \, \mbox{sgn}(y)$, where
$\mbox{sgn}(x)$ is the sign-function. The thin-wall approximation
is valid whenever the wavelength of scattered particles is much
greater than the thickness of the wall. This approximation allows
us to find analytical solutions to the equations of motion and
does not affect the main results of our analysis.
\\
In thin-wall approximation, the solution of the system
(\ref{Eq10}) for $y>0$ is easily found:
\begin{eqnarray}
\label{Eq11}
\alpha_{+}   \!\!& = &\!\! c_1 e^{i p_1 y} + c_2 e^{i
p_2 y}
                         + c_3 e^{-i p_1 y} + c_4 e^{-i p_2 y} \! ,    \nonumber \\
\alpha^*_{-} \!\!& = &\!\! i x_1 c_1 e^{i p_1 y} + i x_2 c_2
                           e^{i p_2 y}
                           - i x_1 c_3 e^{-i p_1 y} \nonumber \\
             \!\!& - &\!\! i x_2 c_4 e^{-i p_2 y} \! ,   \nonumber \\
\delta_{+} \!\!& = &\!\! i x_3 c_1 e^{i p_1 y} + i x_4 c_2
                           e^{i p_2 y} - i x_3 c_3 e^{-i p_1 y}
                           \nonumber \\
           \!\!& - &\!\! i x_4 c_4e^{-i p_2 y} \! ,                \nonumber \\
\delta^*_{-} \!\!& = &\!\! x_5 c_1 e^{i p_1 y} + x_6 c_2
                           e^{i p_2 y} + x_5 c_3 e^{-i p_1 y}
                           \nonumber \\
             \!\!& + &\!\! x_6 c_4 e^{-i p_2 y} \! .
\end{eqnarray}
Here $c_i$ are integration constants,
\begin{equation}
\label{Eq12} p_{1,2} = \sqrt{E^2 - m_{1,2}^2} \; ,
\end{equation}
with $m_{1,2}$ given by Eq.~(\ref{Eq12new}), and
\begin{eqnarray}
\label{Eq13}
x_{1,2} \!\!& = &\!\! \frac{p_{1,2} (E + m_{2,1})}{(E
+ m_L)(E + m_R) - m_D^2} \, ,
\nonumber \\
x_{3,4} \!\!& = &\!\! \frac{p_{1,2} [(E + m_L)(m_R - m_L \pm m_1
\mp m_2) - 2 m_D^2]}
                         {2 m_D [(E + m_L)(E + m_R) - m_D^2]} \, ,
\nonumber \\
x_{5,6} \!\!& = &\!\! \frac{m_R - m_L \pm m_1 \mp m_2}{2 m_D} \, .
\end{eqnarray}
The solution in the case $y<0$ is obtained from Eq.~(\ref{Eq11}),
by the substitutions $c_i \rightarrow d_i$ and $x_i \rightarrow
y_i$, where $d_i$ are new integration constants and
\begin{eqnarray}
\label{Eq14}
y_{1,2} \!\!& = &\!\! \frac{p_{1,2} (E + m_{2,1})}{(E
-m_L)(E - m_R) - m_D^2} \, ,
\nonumber \\
y_{3,4} \!\!& = &\!\! \frac{p_{1,2} [(E - m_L)(m_R - m_L \mp m_1
\pm m_2) + 2 m_D^2]}
                           {2 m_D [(E - m_L)(E - m_R) - m_D^2]} \, ,
\nonumber \\
y_{5,6} \!\!& = &\!\! x_{5,6}.
\end{eqnarray}
Returning to the expression for the chiral spinor fields $\psi_L$
and $\psi_R$, we have
\begin{equation}
\label{Eq15}
\psi_L = \psi_L^{(+)} + \psi_L^{(-)}, \;\;\; \psi_R =
\psi_R^{(+)} + \psi_R^{(-)},
\end{equation}
where $\psi_L^{(\pm)}$ and $\psi_R^{(\pm)}$ are explicitly given
by
\begin{equation}
\label{Eq16}
\psi_L^{(\pm)} = \left( \!\!\!
                       \begin{array}{c}
                            \alpha_{\pm}             \\
                            \alpha^*_{\mp}  \\
                          - \alpha_{\pm}             \\
                          - \alpha^*_{\mp}
                       \end{array}
                 \!\! \right) e^{\mp iEt},  \;\;\;
\psi_R^{(\pm)} = \left( \!\!
                       \begin{array}{c}
                            \delta^*_{\mp}  \\
                            \delta_{\pm}             \\
                            \delta^*_{\mp}  \\
                            \delta_{\pm}
                       \end{array}
                  \!\! \right) e^{\mp iEt}.
\end{equation}
For definiteness we consider solutions for which we have incident
fermion states from the left ($y < 0$) which are scattered into
reflected waves going to the left and transmitted waves going to
the right ($y > 0$). Therefore, the fermions are represented by
incoming and reflected waves to the left of the wall and by
transmitted waves to the right. Hence, taking into account
Eqs.~(\ref{Eq11}) and (\ref{Eq16}) we obtain the transmitted,
incident, and reflected left-handed wave functions:
\begin{eqnarray}
\label{Eq17}
(\psi_L^{(\pm)})^{\mbox{{\scriptsize tran}}} \!\!& =
&\!\! (c_1 u_{L,1}^{(\pm)} \, e^{\pm i p_1 y} +
                             c_2 u_{L,2}^{(\pm)} \, e^{\pm i p_2
                             y})\, e^{\mp i E t},
\nonumber \\
(\psi_L^{(\pm)})^{\mbox{{\scriptsize inc}}} \!\!& = &\!\!
                             (d_1 v_{L,1}^{(\pm)} \, e^{\pm i p_1 y} +
                             d_2 \, v_{L,2}^{(\pm)} \, e^{\pm i p_2
                             y}) \, e^{\mp i
E t},
\nonumber \\
(\psi_L^{(\pm)})^{\mbox{{\scriptsize refl}}} \!\!& = &\!\!  (d_3
v_{L,3}^{(\pm)} \, e^{\mp i p_1 y} +
                             d_4 v_{L,4}^{(\pm)} \, e^{\mp i p_2
                             y}) \, e^{\mp
i E t}, \nonumber \\
\end{eqnarray}
with the condition $c_3 = c_4 = 0$. Here, we have introduced the
spinors
\begin{eqnarray}
\label{Eq18}
(u_{L,1}^{(+)})^{T} \!\!& = &\!\! (1,  i x_1, -1, -i
x_1),
\nonumber \\
(u_{L,2}^{(+)})^{T} \!\!& = &\!\! (1,  i  x_2, -1, -i x_2),
\nonumber \\
(u_{L,3}^{(+)})^{T} \!\!& = &\!\! (1, -i x_1, -1,  i x_1),
\\
(u_{L,4}^{(+)})^{T} \!\!& = &\!\! (1, -i x_2, -1,  i x_2).
\nonumber
\end{eqnarray}
The spinors $v_{L,i}^{(+)}$ are obtained from $u_{L,i}^{(+)}$ by
the replacements $x_i \rightarrow y_i$, while $u_{L,i}^{(-)} = C
u_{L,i}^{(+)}$ and $v_{L,i}^{(-)} = C v_{L,i}^{(+)}$, where $i =
1,2,3,4$, and $C$ is the charge conjugation matrix. The
transmitted, incident, and reflected right-handed wave functions
are obtained from Eq.~(\ref{Eq17}) by the substitutions
$u_{L,i}^{(\pm)} \rightarrow u_{R,i}^{(\pm)}$, $v_{L,i}^{(\pm)}
\rightarrow v_{R,i}^{(\pm)}$, where
\begin{eqnarray}
\label{Eq19}
(u_{R,1}^{(+)})^{T} \!\!& = &\!\! (x_5,  i x_3, x_5 ,
i x_3),
\nonumber \\
(u_{R,2}^{(+)})^{T} \!\!& = &\!\! (x_6,  i x_4, x_6 , i x_4),
\nonumber \\
(u_{R,3}^{(+)})^{T} \!\!& = &\!\! (x_5, -i x_3, x_5 , -i x_3),
\\
(u_{R,4}^{(+)})^{T} \!\!& = &\!\! (x_6, -i x_4, x_6 , -i x_4).
\nonumber
\end{eqnarray}
The spinors $v_{R,i}^{(+)}$ are obtained from $u_{R,i}^{(+)}$ by
the replacements $x_i \rightarrow y_i$, while $u_{R,i}^{(-)} = C
u_{R,i}^{(+)}$ and $v_{R,i}^{(-)} = C v_{R,i}^{(+)}$. By imposing
continuity of $\alpha_{\pm}(y)$ and $\delta_{\pm}(y)$ in $y=0$, we
get
\begin{eqnarray}
\label{Eq20}
c_1 + c_2         \!\!& = &\!\! d_1 + d_2 + d_3 + d_4,                  \nonumber  \\
x_1 c_1 + x_2 c_2 \!\!& = &\!\! y_1 d_1 + y_2 d_2 - y_1 d_3 - y_2 d_4,  \nonumber  \\
x_3 c_1 + x_4 c_2 \!\!& = &\!\! y_3 d_1 + y_4 d_2 - y_3 d_3 - y_4 d_4,             \\
x_5 c_1 + x_6 c_2 \!\!& = &\!\! y_5 d_1 + y_6 d_2 + y_5 d_3 + y_6
d_4.  \nonumber
\end{eqnarray}
Solving the above system with respect to $c_1$, $c_2$, $d_3$,
$d_4$, we obtain:
\begin{eqnarray}
\label{Eq21}
c_1 \!\!& = &\!\! \left (1 + \frac{m_1}{E} \right )
\! d_1, \;\;\;
d_3 = \frac{m_1}{E} \: d_1,                            \nonumber \\
c_2 \!\!& = &\!\! \left (1 + \frac{m_2}{E} \right ) \! d_2, \;\;\;
d_4 = \frac{m_2}{E} \: d_2.
\end{eqnarray}
The total current is defined in terms of the asymptotic fermion
states discussed in Section II:
\begin{eqnarray}
\label{Eq27}
J^{\mu}_{(\pm)}  \!\!& = &\!\!
                          {{\Psi}_M^{(\pm)}}^{\dag} \gamma^0 \gamma^{\mu} \, {\Psi}_M^{(\pm)}
                       = {{\Psi}^{(\pm)}}^{\dag} \gamma^0 \gamma^{\mu} \, {\Psi}^{(\pm)}
                       \nonumber \\
                 \!\!& = &\!\!
                       \left (
                              {\psi_L^{(\pm)}}^{\dag},
                              ({\psi_R^c)^{(\pm)}}^{\dag},
                              ({\psi_L^c)^{(\pm)}}^{\dag},
                              {\psi_R^{(\pm)}}^{\dag}
                       \right )
\nonumber \\
\!\!& \times &\!\!
         \begin{pmatrix}
             \gamma^0 \gamma^{\mu} & 0 & 0 & 0 \\
             0 & \gamma^0 \gamma^{\mu} & 0 & 0 \\
             0 & 0 & \gamma^0 \gamma^{\mu} & 0 \\
             0 & 0 & 0 & \gamma^0 \gamma^{\mu}
         \end{pmatrix} \!\!
                           \left( \!\!\!
                                 \begin{array}{c}
                                    \psi_L^{(\pm)}      \\
                                    (\psi_R^c)^{(\pm)}  \\
                                    (\psi_L^c)^{(\pm)}  \\
                                    \psi_R^{(\pm)}
                                 \end{array}
                      \!\! \right) \! ,
                 \nonumber \\
\end{eqnarray}
where the second equality holds because $U$ is a unitary matrix.
Here ``$\pm$'' refer to positive and negative energy states, and
$\Psi_M^{(\pm)} = U^T \Psi^{(\pm)}$. It should be noted that the
conjugate wave functions are $(\psi_L^c)^{(\pm)} = C
({\bar{\psi}_L^{(\mp)})^T}$ and $(\psi_R^c)^{(\pm)} = C
({\bar{\psi}_R^{(\mp)})^T}$.
\\
Because we are considering the motion of fermions perpendicular to
the wall, the relevant currents are those perpendicular to the
kink, i.e., $J^{2}_{(\pm)}$. Taking into account the expressions
for the chiral wave functions and Eq.~(\ref{Eq27}), we get the
transmitted, incident and reflected currents:
\begin{eqnarray}
\label{Eq28}
{(J^2_{(\pm)})}^{\mbox{{\scriptsize tran}}} \!\! & =
& \!\!
        8 \, [(x_1 + x_3 x_5) c_1^2 + (x_2 + x_4 x_6) c_2^2 \,]
\nonumber \\
         \!\! & = & \!\! \chi^T \mathcal{T} \chi \, ,                   \nonumber \\
{(J^2_{(\pm)})}^{\mbox{{\scriptsize inc}}} \!\! & = & \!\!
        8 \, [(y_1 + y_3 y_5) d_1^2 + (y_2 + y_4 y_6) d_2^2 \,]
\nonumber \\
         \!\! & = & \!\! \chi^T \chi \, ,                               \nonumber \\
{(J^2_{(\pm)})}^{\mbox{{\scriptsize refl}}} \!\! & = & \!\!
         - 8 \, [(y_1 + y_3 y_5) d_3^2 + (y_2 + y_4 y_6) d_4^2 \,]
\nonumber \\
         \!\! & = & \!\! - \chi^T \mathcal{R} \chi \, ,
\end{eqnarray}
where we have introduced the vector
\footnote{It is straightforward to check that the quantities $y_1
+ y_3 y_5$ and $y_2 + y_4 y_6$ are positive definite.}
\begin{equation}
\label{Eq29}
\chi^T = 2 \sqrt{2} \, (\sqrt{y_1 + y_3 y_5} \: d_1
\, ,
                        \sqrt{y_2 + y_4 y_6} \: d_2),
\end{equation}
and the ``reflection and transmission matrices''
\begin{equation}
\label{Eq30}
\mathcal{R} = \begin{pmatrix}
                     m_1^2 / E^2 & 0 \\
                     0 & m_2^2 / E^2
              \end{pmatrix} \!\! ,
\;\;\; \mathcal{T} = \begin{pmatrix}
                     p_1^2 / E^2 & 0 \\
                     0 & p_2^2 / E^2
              \end{pmatrix} \!\! .
\end{equation}
Note that $\mathcal{R} + \mathcal{T} = 1$. For an incident
particle, the reflection and transmission coefficients are given
as the ratios of the corresponding reflected and transmitted
currents. From Eq.~(\ref{Eq28}) we get
\begin{equation}
\label{Eq31}
R = - \frac{{(J^2)}^{\mbox{{\scriptsize refl}}}}
               {{(J^2)}^{\mbox{{\scriptsize inc}}}} \,
  = \frac{\chi^T \mathcal{R} \chi}{\chi^T \chi} \; ,
\;\;\; T = \frac{{(J^2)}^{\mbox{{\scriptsize tran}}}}
              {{(J^2)}^{\mbox{{\scriptsize inc}}}} \;
  = \frac{\chi^T \mathcal{T} \chi}{\chi^T \chi} \, .
\end{equation}
%
%
\begin{figure}[h!]
\begin{center}
\includegraphics[clip,width=0.4\textwidth]{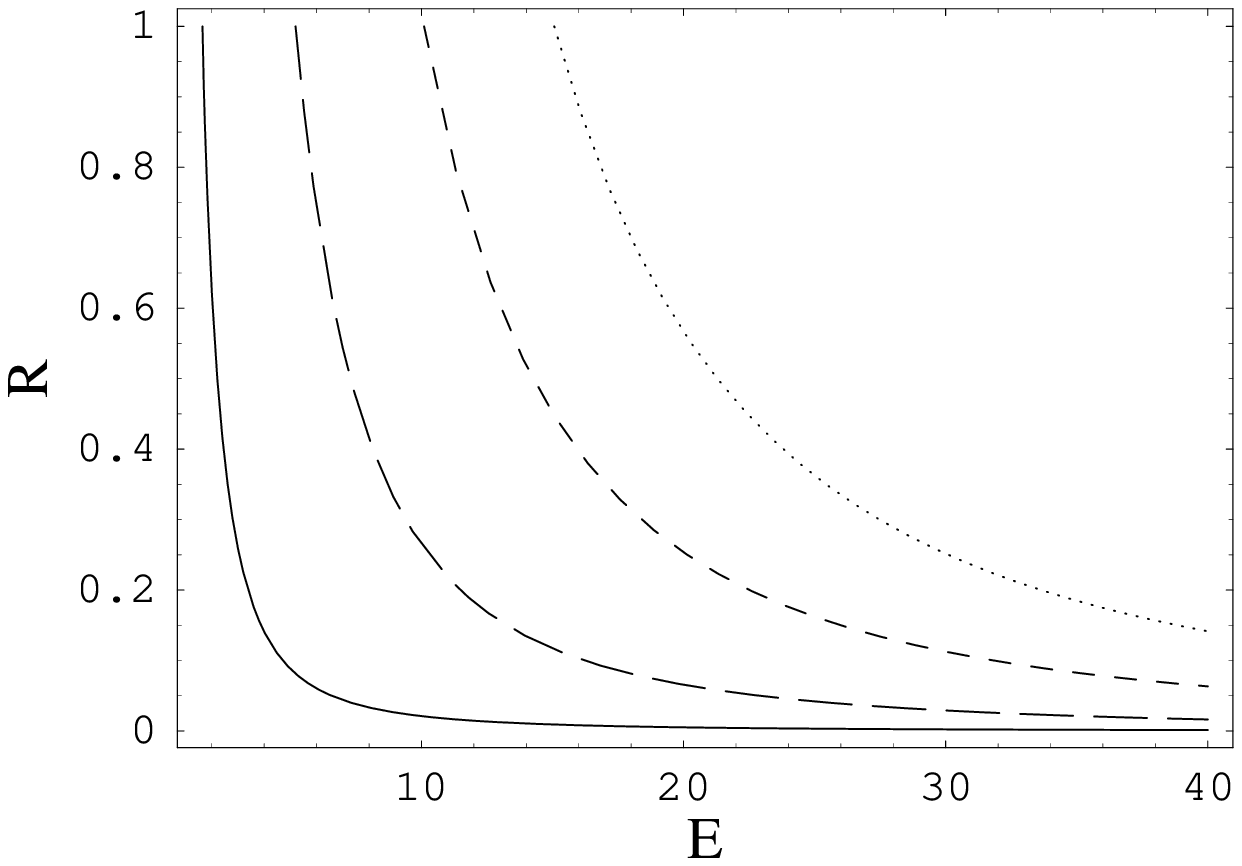}
\includegraphics[clip,width=0.4\textwidth]{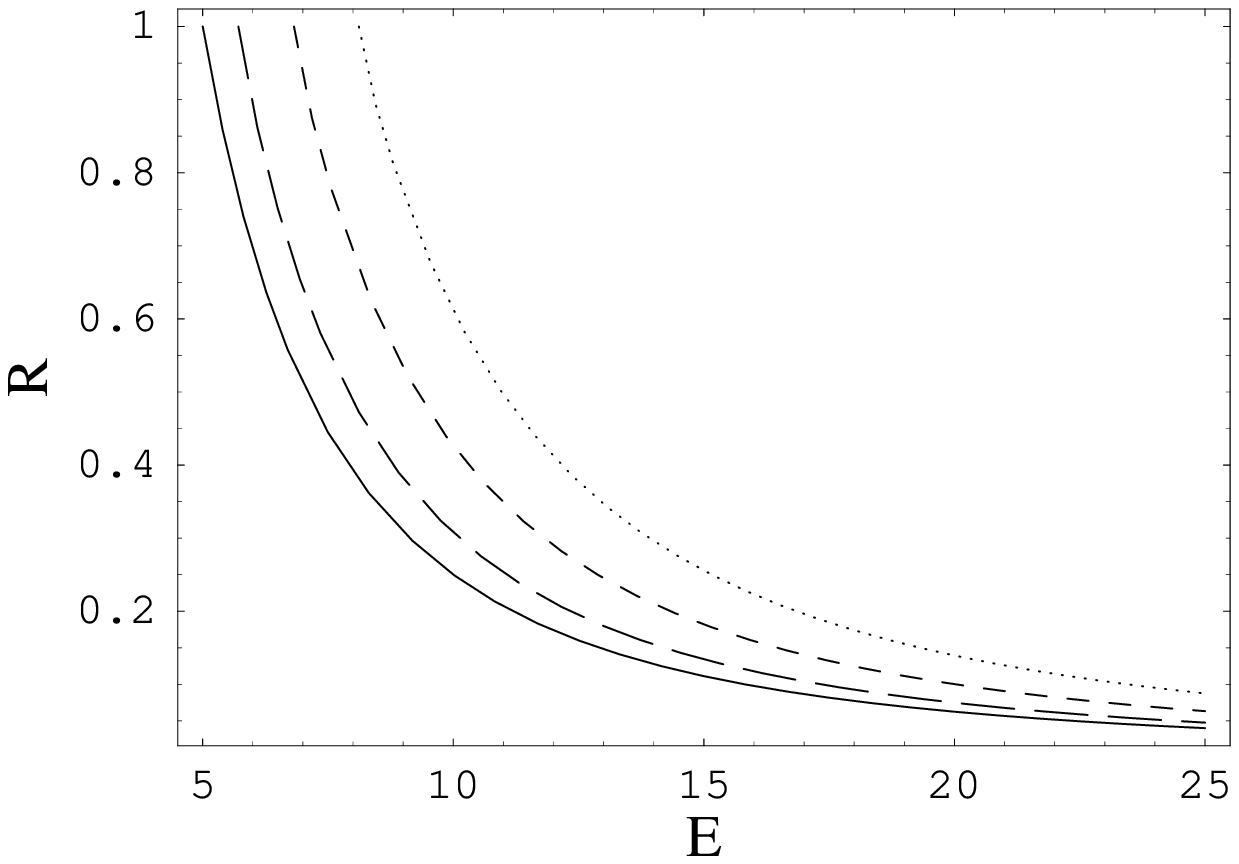}
\includegraphics[clip,width=0.4\textwidth]{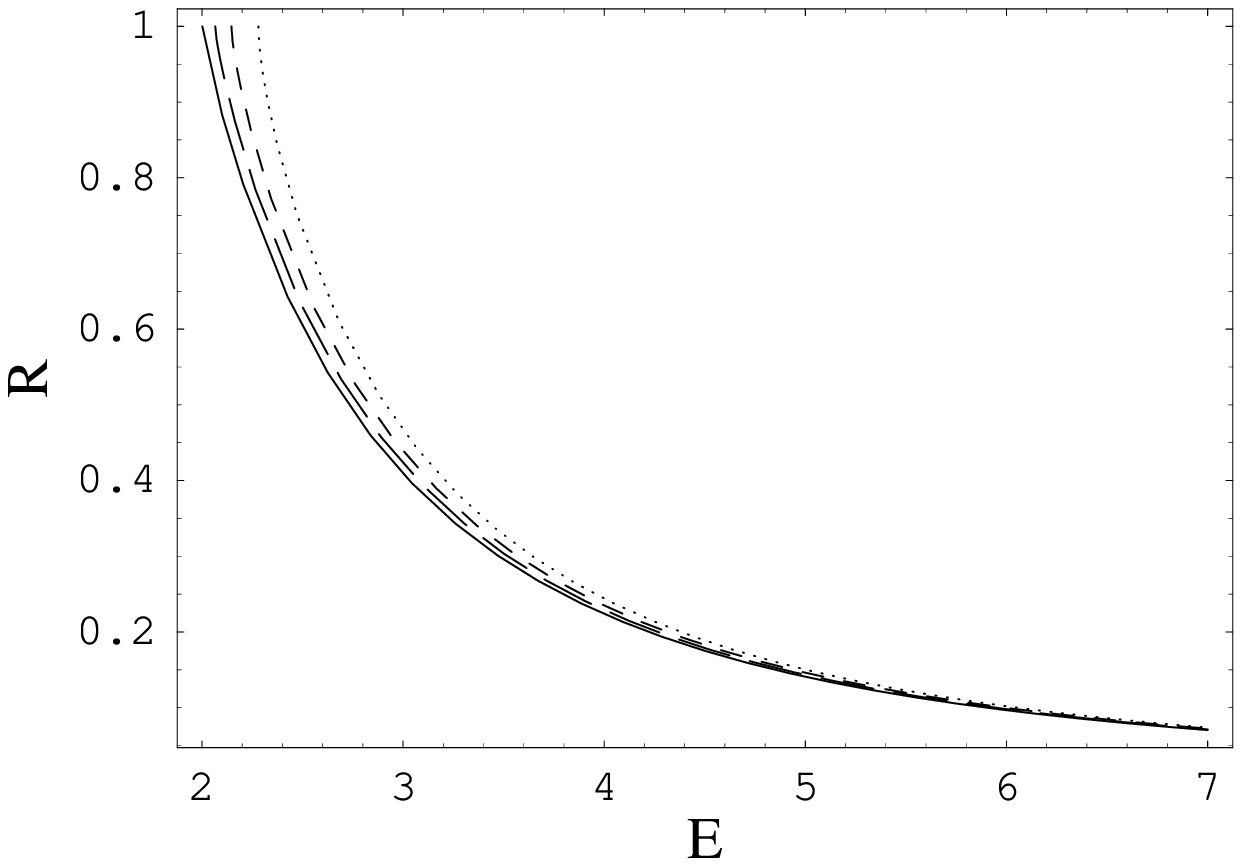}
\caption{Reflection coefficient versus the energy for the case of
non-constant Majorana mass terms, with $d_1 = d_2 = 1/\sqrt{2}$.
{\it Upper panel}. $R$ in the case $m_D = 1$ and $m_L = 0.1$, for
four different values of $m_R$: $m_R = 1$ (solid line), $m_R = 5$
(long-dashed line), $m_R = 10$ (short-dashed line), $m_R = 15$
(dotted line). {\it Middle panel}. $R$ in the case $m_L = 0.1$ and
$m_R = 5$, for four different values of $m_D$: $m_D = 0.1$ (solid
line), $m_D = 2$ (long-dashed line), $m_D = 3.5$ (short-dashed
line), $m_D = 5$ (dotted line). {\it Lower panel}. $R$ in the case
$m_D = 1$ and $m_R = 2$, for four different values of $m_L$: $m_L
= 0.01$ (solid line), $m_L = 0.3$ (long-dashed line), $m_L = 0.6$
(short-dashed line), $m_L = 1$ (dotted line).}
\end{center}
\end{figure}
%
%
Taking into account Eq.~(\ref{Eq30}), the unitary relation, $R + T
= 1$, follows immediately. For $m_L = m_R = 0$, it is
straightforward to check that $R = m_D^2 / E^2$, as it should be
\cite{VILENKIN}.
\\
It should be noted that, since $d_1^2$ and $d_2^2$ are directly
proportional to the amplitudes of the free-field incident wave
functions (i.e. the incident asymptotic fermion states), by a
suitable normalization of wave functions we can take $d_1$ and
$d_2$ such that $d_1^2 + d_2^2 = 1$. Let us observe that the two
incident fermion states of momenta $p_1$ and $p_2$ are scattered
in differen way. Indeed taking $d_1 = 0$ we get $R = m_2^2 / E^2$,
while for $d_2 = 0$ we have $R = m_1^2 / E^2$. We see that the
interaction with vacuum domain walls is able to produce a local
asymmetry in the distribution of the two Majorana fermions states
of masses $m_1$ and $m_2$.

In the upper panel of Fig.~1 we plot the reflection coefficient
versus the energy at fixed $d_1$, $d_2$, $m_D$ and $m_L$, for
different values of $m_R$. In the middle (lower) panel we fix
$m_L$ ($m_D$) and $m_R$, and vary $m_D$ ($m_L$). These figures
show that the reflection coefficient rapidly decreases as the
energy of the incident particles increases, as expected. Indeed,
from Eq.~(\ref{Eq31}) we get that $R \simeq A / E^2$ for $E \gg
m_{1,2}$, where $A$ is a constant depending on $d_1$, $d_2$,
$m_1$, and $m_2$. If $m_L \ll m_D \ll m_R$ and $d_1 = d_2$, then
$A = m_2^2$. Moreover, at fixed energy, fixing two of the three
masses $m_D$, $m_L$, $m_R$, the reflection coefficient is an
increasing function of the remaining mass parameter
\footnote{Because the most relevant phenomenological model for
neutrino masses is the so-called ``see-saw mechanism''
\cite{Grossman} in which $m_L = 0$ and $m_D \ll m_R$, in our
figures we have taken $m_L \leq m_D \leq m_R$.}.
The essential properties of $R$ above discussed does not change if
we take $d_1 \neq d_2$. Indeed, the only effect to take $d_1 >
d_2$ ($d_1 < d_2$) is, at fixed energy and mass parameters, to
increase (decrease) the reflection coefficient.

As pointed out in Ref.~\cite{Stojkovic}, the Majorana masses could
arise from the coupling to a scalar field which undergoes a phase
transition above the phase transition of the field $\Phi$. In this
case, the source of the Majorana masses is not the coupling with
$\Phi$, and the Majorana mass terms are spatially homogeneous. In
this case we set
\begin{equation}
\label{Eq32}
g_L \Phi \rightarrow m_L, \;\;\; g_R \Phi \rightarrow
m_R,
\end{equation}
in Lagrangian (\ref{Eq3}). In thin-wall approximation, the
solution of the equations of motion is, for $y>0$, equal to
Eq.~(\ref{Eq11}), while for $y<0$ is:
\begin{eqnarray}
\label{Eq33}
\alpha_{+}   \!\!& = &\!\! d_1 e^{i p_1 y} + d_2 e^{i
p_2 y}
                   + d_3 e^{-i p_1 y} + d_4 e^{-i p_2 y},  \nonumber \\
\alpha^*_{-} \!\!& = &\!\! i x_1 d_1 e^{i p_1 y} + i x_2 d_2
e^{ip_2 y}
- i x_1 d_3 e^{-i p_1 y} \nonumber \\
                           \!\!& - &\!\! i x_2 d_4 e^{-i p_2 y}, \nonumber \\
\delta_{+} \!\!& = &\!\! - i x_3 d_1 e^{i p_1 y} - i x_4 d_2
                           e^{i p_2 y}
           + i x_3 d_3 e^{-i p_1 y} \nonumber \\
                           \!\!& + &\!\! i x_4 d_4 e^{-i p_2 y}, \nonumber  \\
\delta^*_{-} \!\!& = &\!\! - x_5 d_1 e^{i p_1 y} - x_6 d_2
                           e^{i p_2 y}
              - x_5 d_3 e^{-i p_1 y} \nonumber \\
                           \!\!& - &\!\! x_6 d_4 e^{-i p_2 y},
\end{eqnarray}
where $d_i$ are constants of integration, $p_{1,2}$ and $m_{1,2}$
are the same as in Eq.~(\ref{Eq12}), and $x_i$ are given by
Eq.~(\ref{Eq13}). Taking into account the expressions for
$\psi_{L}^{(\pm)}$ and $\psi_{R}^{(\pm)}$ (see Eq.~(\ref{Eq16})),
and Eq.~(\ref{Eq33}), we obtain the transmitted, incident, and
reflected left-handed wave functions:
\begin{eqnarray}
\label{Eq34}
(\psi_L^{(\pm)})^{\mbox{{\scriptsize tran}}} \!\!& =
&\!\! (c_1 u_{L,1}^{(\pm)} \, e^{\pm i p_1 y} + c_2
u_{L,2}^{(\pm)} \, e^{\pm i p_2 y}) \, e^{\mp i E t},
\nonumber \\
(\psi_L^{(\pm)})^{\mbox{{\scriptsize inc}}} \!\!& = &\!\!
                             (d_1 u_{L,1}^{(\pm)} \, e^{\pm i p_1 y} +
                             d_2 u_{L,2}^{(\pm)} \, e^{\pm i p_2
                             y}) \, e^{\mp i E t},
\nonumber \\
(\psi_L^{(\pm)})^{\mbox{{\scriptsize refl}}} \!\!& = &\!\!  (d_3
u_{L,3}^{(\pm)} \, e^{\mp i p_1 y} +
                             d_4 u_{L,4}^{(\pm)} \, e^{\mp i p_2
                             y}) \, e^{\mp i E t},
\nonumber \\
\end{eqnarray}
where the spinors $u_{L,i}^{(\pm)}$ are given by Eq.~(\ref{Eq18}).
The transmitted, incident, and reflected right-handed wave
functions are obtained from Eq.~(\ref{Eq34}) by the substitutions
$u_{L,i}^{(\pm)} \rightarrow u_{R,i}^{(\pm)}$, and $d_i
\rightarrow -d_i$, where the spinors $u_{R,i}^{(\pm)}$ are given
by Eq.~(\ref{Eq19}).
\\
Taking into account Eqs.~(\ref{Eq27}) and (\ref{Eq34}), we get the
transmitted, incident and reflected currents:
\begin{eqnarray}
\label{Eq35}
{(J^2_{(\pm)})}^{\mbox{{\scriptsize tran}}} \!\! & =
& \!\!
       8 \, [(x_1 + x_3 x_5) c_1^2 + (x_2 + x_4 x_6) c_2^2 \,] ,
                                                                 \nonumber \\
{(J^2_{(\pm)})}^{\mbox{{\scriptsize inc}}} \!\! & = & \!\!
       8 \, [(x_1 + x_3 x_5) d_1^2 + (x_2 + x_4 x_6) d_2^2 \,] ,
                                                                           \\
{(J^2_{(\pm)})}^{\mbox{{\scriptsize refl}}} \!\! & = & \!\!
         - 8 \, [(x_1 + x_3 x_5) d_3^2 + (x_2 + x_4 x_6) d_4^2 \,] .
                                                                 \nonumber
\end{eqnarray}
Now, imposing continuity of $\alpha_{\pm}(y)$ and
$\delta_{\pm}(y)$ in $y=0$, we get
\begin{eqnarray}
\label{Eq36}
c_1 + c_2         \!\!& = &\!\!   d_1 + d_2 + d_3 + d_4,                  \nonumber  \\
x_1 c_1 + x_2 c_2 \!\!& = &\!\!   x_1 d_1 + x_2 d_2 - x_1 d_3 - x_2 d_4,  \nonumber  \\
x_3 c_1 + x_4 c_2 \!\!& = &\!\! - x_3 d_1 - x_4 d_2 + x_3 d_3 + x_4 d_4,             \\
x_5 c_1 + x_6 c_2 \!\!& = &\!\! - x_5 d_1 - x_6 d_2 - x_5 d_3 -
x_6 d_4. \nonumber
\end{eqnarray}
Solving the above system with respect to $c_1, c_2, d_3, d_4$, and
inserting the solution into Eq.~(\ref{Eq35}), we obtain, after
some manipulations, the reflection and transmission coefficients:
\begin{eqnarray}
\label{Eq37}
R \!\!& = &\!\! \frac{2 m_D^2 E^2}{2 m_D^2 E^2 + p_1
p_2 (E^2 + p_1 p_2 - m_1 m_2 )} \; ,
\nonumber \\
T \!\!& = &\!\! \frac{p_1 p_2 (E^2 + p_1 p_2 - m_1 m_2 )}
               {2 m_D^2 E^2 + p_1 p_2 (E^2 + p_1 p_2 - m_1 m_2 )} \; .
\end{eqnarray}
The unitary condition follows immediately from Eq.~(37) and for
$m_L = m_R = 0$ we get $R = m_D^2 / E^2$, as it should be
\cite{VILENKIN}.
\\
It is interesting to note that in the case of spatially
homogeneous Majorana mass terms the reflection and transmission
coefficients do not depend on the amplitudes of the two incident
asymptotic fermion states of momenta $p_1$ and $p_2$. In fact, the
dependence due to the amplitudes of this states factorizes in the
expression of the currents in such a way that the reflection and
transmission coefficients do not show any explicit dependence on
$d_1$ and $d_2$. Since the two Majorana fermions states of masses
$m_1$ and $m_2$ are scattered in the same way, there is no
production of local asymmetry of any kind.

The behavior of $R$ as a function of one of the tree mass
parameters (keeping constant the other two) is the same as in the
case of non-constant Majorana mass terms, while for large values
of energy, $E \gg m_{1,2}$, the reflection coefficient decreases
as $R \simeq m_D^2 / E^2$.


\section{IV. Conclusions}
We studied the interaction of fermions having both Dirac and
left-handed and right-handed Majorana mass terms with kink domain
walls. The source of the Dirac mass term was taken to be the
coupling to the scalar field $\Phi$ that gives rise to a wall. As
regards the source of the Majorana mass terms, we analyzed two
possible cases. In the first case we assumed that the Majorana
masses are generated by the coupling to the scalar field $\Phi$,
in the second one, the Majorana mass terms were taken to be
spatially homogeneous.
\\
We found the asymptotic fermion states with definite masses, $m_1$
and $m_2$, which represent the free-field propagating degrees of
freedom in the theory.

By solving the Dirac equation in thin-wall approximation, we
calculated the reflection and transmission coefficients for the
scattering of such fermions off walls. The peculiar properties of
the reflection coefficient $R$ were analyzed in both cases of
non-constant and constant Majorana mass terms.
\\
In the case of non-constant Majorana mass terms, the fermion
states with definite masses scatter with different probabilities.
Indeed, if the incident state consist of a state of definite mass
$m_1$ or $m_2$, then his scattering probability is $R = m_1^2 /
E^2$ or $R = m_2^2 / E^2$, respectively. In the case in which the
incident state is a superposition of the two definite mass states,
then the reflection coefficient has a quite complicated
expression. However, for high energy of the incident particles, it
is given by $R \simeq m_2^2 / E^2$ (in the limit $m_L \ll m_D \ll
m_R$).
\\
In the case of constant Majorana mass terms, the fermion states
with definite masses scatter in the same way. We found that for
high energy of the incident particles, the reflection coefficient
is $R \simeq m_D^2 / E^2$.

We conclude by stressing that the reflection coefficients we found
in this paper are important for determining the equation of state
of a gas of domain walls which could be present in our Universe.
The scattering of Majorana particles off vacuum domain walls,
together with the presence of localized zero modes, could strongly
influence the cosmic evolution of a gas of domain walls. However,
the discussion of this last issue is beyond the aim of this paper
and will be the object of future investigations.

\vspace*{0.5cm}


\begin{acknowledgments}
I would like to thank A. Marrone for reading the manuscript and
for helpful comments. I also thank A. Palazzo and M. Ruggieri for
useful discussions, and an anonymous referee for helpful comments
and suggestions that greatly improved the paper.
\end{acknowledgments}



\end{document}